\documentstyle[12pt]{article}                 
\title{ New stochastic approach to the renormalization of the supersymmetric 
$\phi^4$ with ultrametric.}      
	         
\author{Suemi Rodr\'\i guez-Romo\\Centre of Theoretical Research\\
National University of M\'exico, Campus Cuautitl\'an\\ 
Apdo. 95 Unidad Militar, Cuautitl\'an Izcalli\\ 
Edo. de M\'exico, 54768 M\'exico. $^{\ast}$}         
     
\date{}  
  
\begin{document}

\maketitle

\renewcommand{\thefootnote}{\fnsymbol{footnote}}
\setcounter{footnote}{-1}
\footnote{$\hspace*{-6mm}^{\ast}$
e-mail: suemi$@$servidor.unam.mx \\
$\hspace*{1.5cm}$ suemi@fis.cinvestav.mx}
\renewcommand{\thefootnote}{\arabic{footnote}}\

\baselineskip0.4cm

{\small {\bf Abstract.} We present a new real space renormalization-group 
map, on the space of probabilities, to study the renormalization of the SUSY 
$\phi^4$. In our approach we use the random walk representation on a lattice 
labeled by an ultrametric space. Our method can be extended to any 
$\phi^n$. New stochastic meaning is given to the parameters involved in the 
flow of the map and results are provided.\\

PACS 8.03.65.Bz}

\newpage

\baselineskip0.6cm  

\section{Introduction}

In this paper we use a new real space renormalization-group map to 
study renormalization of SUSY $\phi^4$ theories. 
Symanzik's \cite{Sy} work proved that $\phi^4$ theories can be 
represented as weakly self-avoiding random walk models. They are 
in the same universality class as the self-avoiding random walk \cite{La}.\ 

Our renormalization-group transformation is carried out  on the space of 
probabilities. Real space renormalization-group methods have proved to be 
useful in the study of a wide class of phenomena and are used here
to provide a new stochastic meaning to the parameters involved in the flow of
the interacting constant and the mass, as well as the $\beta$ function for
SUSY $\phi^4$.\

The hierarchical models introduced by Dyson \cite{Dy} have the feature of 
having a simple renormalization-group transformation. We use a hierarchical 
lattice where the points are labeled by elements of a countable, abelian 
group {\it G} with ultrametric $\delta$; i.e. the metric space 
$({\it G},\delta)$ is hierarchical. The hierarchical structure of this metric space induces a renormalization-group 
map that is ``local"; i.e. instead of studying the space of random functions 
on the whole lattice, we can descend to the study of random functions on 
L-blocks (cosets of {\it G}) \cite{Ev}. Our method provides a probabilistic 
meaning to every parameter appearing in the flow of interaction constant  and 
the mass for any $\phi^n$.\ 

This paper is organized as follows; in Section 2 we present
the lattice and the corresponding class of L\'{e}vy processes which are
studied here. In Section 3 we define the renormalization-group map and apply 
this to  SUSY $\phi^4$, in random walk representation. In Section 4 we use 
the results obtained in previous Section to give new ligth into the stochastic 
meaning of the map and results are presented.\

\section{The lattice with ultrametric and the L\'{e}vy process.}

The hierarchical lattice used in this paper was introduced by Brydges, Evans
and Imbrie \cite{Ev}.Here, we present a slight variant. Fix an integer 
$L\geq 2$. The points of the lattice are labeled by elements of the countable, 
abelian group ${\it G}=\oplus ^{\infty}_{k=0}{\bf Z}_{L^d}$, d being the 
dimension of the lattice.  A one-dimensional example can be found in Brydges 
et al \cite{Ev} and a two dimensional example in Rodr\'{\i}guez-Romo \cite{Su}. 
An element $X_i$ in ${\it G}$ is an infinite sequence
$$
X_i\equiv (...,y_k,...,y_2, y_1, y_0 )\;\;;\;\;y_i\in {\bf Z}_{L^d}\;\;
\mbox{  thus  }\;\; X_i\in {\it G}=\oplus^{\infty}_{k=0}{\bf Z}_{L^d},
$$
where only finitely many $y_i$ are non-zero.\

Let us define subgroups
\begin{equation}
\{0\}={\it G}_0\subset {\it G}_1\subset ...\subset {\it G}
\;\;\;\mbox{where  }
{\it G}_k=\{X_i\in {\it G}| y_i=0, i\geq k \}
\end{equation}
and the norm $|\cdot|$ as
\begin{equation}
|X_i|=\left\{\begin{array}{cc}
0 & \mbox{ if $X_i=0$} \\
L^p & \mbox{where   $p=\inf\{k| X_i\in {\it G}_k\}$  if   $X_i\neq 0$}
\end{array}
\right.
\end{equation}
Then, the map $\delta:(X_{i+1}, X_i)\rightarrow |X_{i+1}-X_i|$ defines a 
metric on ${\it G}$. In this metric the subgroups ${\it G}_k$ are balls 
$|X_i|\leq L^k$ containing $L^{dk}$ points. Here the operation + (hence - as 
well) is defined componentwise.\

The metric defined by eq(2) satisfies a stronger condition than the triangle 
inequality, i.e.
\begin{equation}
|X_i+X_{i+1}|\leq \mbox{ Max}(|X_i|,|X_{i+1}|).
\end{equation}
From eq(3), it is clear that  the metric introduced is an ultrametric. \

For the purposes of this paper we introduce the L\'{e}vy process as a 
continuous time random walk $w$. This is the following ordered sequence of 
sites in {\it G};
\begin{equation}
(w(t_0),...,w(t_0+...+t_n))\;\;,w(t_0+...+t_i)=X_i\in {\it G},
\;\;T=\sum ^n_{i=0}t_i ,\; n\geq 0
\end{equation}
where $t_i$ is the time spent in $X_i\in {\it G}$ (waiting time at $X_i$)
and T, fixed at this point, is the runnig time for the process. For convenience we take 
$X_0=0$.\ 

We are not dealing with nearest neighbour random walks on the lattice, 
provided we mean neighbourhood with respect to the ultrametric distance 
$\delta$ previously defined. We propose the L\'{e}vy process we are dealing 
with, having a probability $P(w)=r^ne^{-rT}\prod^{n-1}_{i=0}q(X_{i+1},X_{i})$. 
Namely the continuous time random walk has a probability $rdt$ (r is the 
jumping rate) of making a step in time $(t,dt)$ and, given that it jumps, the 
probability  of jumping from $X_i$ to $X_{i+1}$ is $q(X_{i+1},X_i)$, 
conditioned to a fixed running time $T$ for the process. $q(X_{i+1},X_i)$ is 
an open function of the initial and final sites of jumps in the lattice 
${\it G}$. Here we define $Dw$ by $\int(\cdot)Dw$=\newline
$\sum_n\sum_{[X_i]^n_{i=0}}\int^T_0\prod^n_{i=0}dt_i
\delta(\sum^n_{i=0}t_i-T)(\cdot)$. From this follows $\int P(w)Dw$=$1$.\

Let the space of simple random walks of length $n$, be $\Lambda_n$, with
probability measure $P(w)$, we construct on this space the weakly SARW model 
that represents SUSY $\phi^4$ (through McKane-Parisi-Sourlas theorem 
\cite{MKPS}). We take advantage of this feature to provide a better 
understanding of SUSY $\phi^4$ renormalization in terms of stochastic 
processes. This method can be straightforward generalized to any SUSY 
$\phi^n$ with ultrametric.

\section{The renormalization-group map on the random walk representation of
SUSY $\phi^4$.}   

We propose a renormalization-group map on the lattice $R(X_i)=LX'_i$ where 
$X_i\in {\it G}$ and $LX'_i\in {\it G}'$=${\it G}/{\it G}_1\sim {\it G}$; i.e.
the renormalized lattice ${\it G}'$ is isomorphic to the original lattice 
${\it G}$. Here $LX'_i=(...,y_2,y_1)$.\ 

Besides we propose the action of the renormalization-group map on the space of 
random walks $R(w)=w'$, from $w$ 
above as defined, to $w'$. Here, 
$w'$ is the following ordered sequence of sites in ${\it G}'=
{\it G}/{\it G}_1\approx {\it G}$; 
\begin{equation}
(w'(t'_0),...,w'(t'_0+...+t'_k))\;\;,\mbox{ where}
\end{equation}
$$
w'(t'_0,...,t'_{i'})=X'_{i'}\in {\it G},\;\;T'=\sum ^k_{i'=0}t'_i ,
\;0\leq k\leq n,
\;\;T=L^{\varphi}T'.
$$
$R$ maps $w(t_0)+{\it G}_1,w(t_0+t_1)+{\it G}_1,...,w(t_0+...+t_n)+{\it G}_1$
to cosets $Lw(t_0)$,$Lw(t_0+t_1)$,...,$Lw(t_0+...+t_n)$ respectively. If two 
or more successive cosets in the image are the same, they are listed only as one 
site in $w'(t'_0),...,w(t'_0+...+t'_k)$, and the times $t'_j$ are sums of the 
corresponding $t_i$ for which successive cosets are the same, rescaled by 
$L^{\varphi}$. For $\varphi=2$, we are dealing with normal diffusion (this is
the standard version of SUSY $\phi^4$), in case $\varphi<2$ with superdiffusion,
and subdiffusion for $\varphi>2$. In the following this parameter is arbitrary,
so we can study general cases.\

We can now work out probability measures at the $(p+1)^{th}$ stage in the
renormalization provided only that we know the probabilities at the $p^{th}$
stage. We integrate the probabilities of all the paths $w^{(p)}$ 
consistent with a fixed path $w^{(p+1)}$ in accordance with the following. 
Let $R(w)=w'$ be the renormalization-group map above as stated, then
$P'(w')$= \newline 
$L^{\varphi k}\int Dw P(w)\chi (R(w)=w')$. Here $R(w)=w'$ is a 
renormalization-group transformation that maps an environment $P(w)$ to a new 
one, $P'(w')$, thereby implementing the scaling.\

Hereafter 
\begin{equation}
m_{j'}=\sum^{j'}_{i'=0}n_{i'}+j' \;\;\mbox{ and}
\end{equation}
$$
n=\sum^{k}_{i'=0}n_{i'}+k\;\;\;\;\;\;0\leq j'\leq k, \mbox{ being}
$$
$n_{i'}$=$max\{i|w(t_0+...+t_j)\in LX_{i'}, \forall j\leq i\}$; i.e. the 
number of steps (for paths $w$) in the contracting ${\it G}_1$ coset that, once 
the renormalization-group map is applied, has the image $LX'_{i'}$.\

Concretely $P'(w')$ can be written like  
\begin{equation}
P'(w')=
L^{\varphi k}\;\sum_{\left[ n_{i'}\right]^{k}_{i'=0}}
\;\sum_{\left[ X_{i}\right]^{n}_{i=0}}
\int \prod^{n}_{i=0}dt_{i}\;
\prod^{k}_{j'=0}\;
\delta (\sum^{m_{j'}}_{{i}=m_{j'-1}+1}\;\;t_{i}-
L^{\varphi }t'_{j'})\times
\end{equation}
$$
\times
\prod^{k}_{j'=0}\;
\prod^{m_{j'}}_{{i}=m_{j'-1}+1}\;
\chi (X_{i}\in LX'_{j'}\;\;)P(w).
$$
It is straightforward to prove that the probability $P(w)$ where we substitute
$q(X_{i+1},X_i)$ by $c|X_{i+1}-X_i|^{-\alpha}$ $\forall X_i,X_{i+1}\in {\it G}$ 
($c$ is a constant fixed up to normalization and $\alpha$ another constant), 
is a fixed point of the renormalization-group map provided $\varphi$=
$\alpha-d$. Even more, if in $P(w)$ we substitute $q(X_{i+1},X_i)$ by 
$c\left(|X_{i+1}-X_i|^{-\alpha}+|X_{i+1}-X_i|^{-\gamma}\right)$ 
$\forall X_i,X_{i+1}\in {\it G}$, $\gamma>>\alpha$ ($\gamma$ is an additional constant), 
then this flows to the very same fixed point of the renormalization-group 
map that in the first case. This holds provided 
$log\left(\frac{L^{-\alpha}-L^{-\gamma}-2L^{d-\gamma-\alpha}}
{L^{-\alpha}-2L^{d-\gamma}}\right)\rightarrow 0$ and $\varphi$=$\alpha-d$.\

Let us substitute $q(X_{i+1},X_i)$ by 
$q_1(|X_{i+1}-X_i|)+\epsilon b(X_{i+1},X_i)$ where $q_1(|X_{i+1}-X_i|)$ is any
function of the distance between $X_{i+1}$ and $X_i$, both sites in the
lattice; $b(X_{i+1},X_i)$ is a random function and $\epsilon$ a small 
parameter. We can impose on $b(X_{i+1},X_i)$ the following conditions.\

a) $\sum_{X_{i+1}}b(X_{i+1},X_i)$=$0$.\

b) Independence. We take $b(X_{i+1},X_i)$ and $b(X'_{i+1},X'_i)$ to be 
independent if $X_{i+1}\neq X'_i$.\

c) Isotropy.\

d) Weak randomness.\

In this case, $P(w)$ is still a fixed point of the renormalization-group map
provided\newline
\begin{equation}
\sum_{(n_j)^k_0}\sum_{(X_i)^n_0}\prod^{n-1}_{i=0}b(X_{i+1},X_i)
\prod^k_{j=0}\prod^{m_j}_{i=m_{j-1}+1}\chi(X_i\in LX'_j)=
\end{equation}
$$
\prod^k_{j=0}b(X'_{j+1},X'_j)L^{-\varphi k}(b(1)(L^d-1))^{n_j}
$$
where $b(1)=\frac{1-L^{-\varphi}}{L^d-1}$ is the probability of jumping from
one specific site to another specific site inside the ${\it G}_1$ coset.\

One formal solution to eq(8) is the following
\begin{equation}
b(X_{i+1},X_i)=\left\{\begin{array}{cc}
\frac{1-L^{\varphi}}{L^d-1} & \mbox{ if $|X_{i+1}-X_i|=L$} \\
\sum_t\left(
\begin{array}c
d+\varphi\\
t
\end{array}
\right)
f(X_{i+1})^tf(X_i)^{d+\varphi-t} & \mbox{where $|X_{i+1}-X_i|>L$}
\end{array}
\right.
\end{equation}
up to proper normalization. Here $f(X_i)$ and $f(X_{i+1})$ are homogeneous
function of sites in the lattice, order -1. Besides they add to $...,1)$ and
are positive defined. Since in the limit $d+\varphi\rightarrow \infty$ (provided 
the mean remains finite) binomial probability distribution tends to Poissson 
distribution; we think a nontrivial SUSY $\phi^4$ theory could be included in 
this case \cite{Kl}.\

The random walk representation of the SUSY $\phi^4$ is a weakly SARW 
that penalizes two-body interactions, this is a configurational measure model.
Configurational measures are measures on $\Lambda_n$. Let $P_U(w)$ be the 
probability on this space such that
\begin{equation}
P_U(w)=\frac{U(w)P(w)}{Z}
\end{equation}
where $Z=\int U(w)P(w)Dw$ and $P(w)$ is the probability above described, thus 
a fixed point of the renormalization-group map. Besides $U(w)$ is the energy 
of the walks. To study the effect of the renormalization-group map on $P_U(w)$ we need to follow the trajectory of $U(w)$ after applying 
several times the renormalization-group map.\

Therefore, from previous definition of the renormalization-group map follows;
$$
P'_{U'}(w')=L^{\varphi k}\int  
P_{U}(w)
\chi (R(w)=w')Dw
$$
where $Z'=Z$, thus
\begin{equation}
U'(w')=
\frac{\int Dw P(w)\chi (R(w)=w')U(w)}
{\int Dw P(w)\chi (R(w)=w')}
\end{equation}
Note that eq(11) can be view as the conditional expected value of $U(w)$
given that the renormalization-group map is imposed. Therefore and hereafter, 
to simplify notation, we write eq(11) as $U'(w')=< U(w) >_{w'}$.\

In the random walk representation of the SUSY $\phi^4$ model with 
interaction $\lambda$, and mass (killing rate in the stochastic framework) 
$m$, $U$ is as follows. 
\begin{equation}
U(w)=\prod_{X\in {\it G}}
e^{-m\sum_{i\in J_{X}}t_{i}-\lambda \sum_{i<j\in J_{X}}
t_{i}t_{j}
{\bf 1}_{\left\{w(t_i)=w(t_j)\right\} }},
\end{equation}
being $m<0$ and $\lambda\stackrel{>}{\scriptscriptstyle<}0$ (small) constants.
Here we set a randomly free running time for the process $T$. The probability 
$P_U(w)$, where $U(w)$ is defined as in eq(12), flows to a fixed form after 
the renormalization-group map is applied. This fixed form is characterized by 
the renormalized  energy
\begin{equation}
U'(w')=
\prod_{X'\in{\it G}}
e^{
-m'
\sum_{i'\in J_{X'}}
t'_{i'}-
\lambda'
\sum_
{\stackrel{i'< j'}
{\left\{i',j'\right\}\in J_{X'}}}
t'_{i'}t'_{j'}
{\bf 1}_{(w(t'_{i'})=w(t'_{j'}))}} 
\times 
\end{equation}
$$
\left\{1+\eta'_{1}
\sum_{
\stackrel{i'< j'< k'}
{\left\{i',j',
k'\right\}\in J_{X'}}}
t'_{i'}t'_{j'}
t'_{k'}
{\bf 1}_{(w(t'_{i'})=w(t'_{j'})
=w(t'_{k'}))}+
\right.
$$
$$
+\left.
\eta'_2
\sum_
{\stackrel{i'< j'}
{\left\{i',j'\right\}\in J_{X'}}}
t'_{i'}t'_{j'}
{\bf 1}_{(w(t'_{i'})=w(t'_{j'}))} 
+\eta'_{3}\sum_{i'\in J_{X'}}
t'_{i'}\right\}+r'.
$$
Here  
\begin{equation}
m' = L^{\varphi}m+m'_1 \mbox{$\;\;\;where$}
\end{equation}
\begin{equation}
m'_{1}= \gamma_{1}\lambda-\gamma_{2}\lambda^2+r_{m'_1}.
\end{equation}
\begin{equation}
\lambda'=L^{2\varphi-d}\lambda-\chi\lambda^2+r_{\lambda'}.
\end{equation}
\begin{equation}
\eta'_{1}=\eta_{1}L^{3\varphi-2d}+\eta \lambda^2.
\end{equation}
\begin{equation}
\eta'_2=\eta_1A+L^{(2\varphi-d)}\eta_2 \mbox{$\;\;\;and$}
\end{equation}
\begin{equation}
\eta'_3=\eta_1B+\eta_2\gamma_{1}+L^{\varphi}\eta_{3}.
\end{equation}

All parameters involved in eq(15), eq(16), eq(17), eq(18) and eq(19);
namely $\gamma_1$, $\gamma_2$, $\chi$, $\eta$, $A$ and $B$ have precise, well 
defined formulae \cite{Su}. They are linearized conditional expectations of 
events inside contracting ${\it G}_1$ cosets which, upon renormalization, maps 
to a fixed random walk with totally arbitrary topology. Even more, we have 
precise formulae for all the remainders, also \cite{Su}. Concretely speaking,
$\gamma_1$ and $\gamma_2$ are contributions to renormalized local times
coming from one and two two-body interactions inside the contracting 
${\it G}_1$ cosets, respectively. $\chi$ is the two two-body interaction 
(inside the contracting ${\it G}_1$ cosets) contribution to renormalized 
two-body interaction. $\eta$ is the contribution to renormalized three-body 
interaction coming from two two-body interactions. Finally $A$ and $B$ are the 
one three-body interaction (inside the contracting ${\it G}_1$ cosets) 
contribution to renormalized two-body interaction and local time, 
respectively.\

In the SUSY representation of $\phi^4$ we can say that $\gamma_1$ 
and $\gamma_2$ are first and second order contributions of SUSY $\phi^4$ to 
renormalized SUSY mass; $\chi$ is the second order contribution of SUSY 
$\phi^4$ to renormalized SUSY $\phi^4$; $\eta$ is the second order contribution 
(the first order contribution is null due to topological restriccions) of SUSY 
$\phi^4$ to renormalized SUSY $\phi^6$. Finally, $A$ and $B$ are first order 
contributions of SUSY $\phi^6$ (already generated at this stage by previous 
renormalization stages) to renormalized SUSY $\phi^4$ and mass, respectively.\

Eq(13) is presented in terms of the product of two factors. The first one
(exponential) involves only; a) trivial flow of mass and interacting constant
b) $\lambda\phi^4$ contribution (inside contracting ${\it G}_1$ cosets) to
renormalized mass and $\lambda'\phi^4$ up to leading order. The second factor 
involves mixed terms; namely $\lambda\phi^4$ and $\phi^6$ contributions 
(inside the contracting ${\it G}_1$ cosets) to renormalized mass, $\phi^4$ and
$\phi^6$. $\phi^6$ terms come into the scheme because they are produced from 
$\lambda\phi^4$ due to the fixed topology of the continuous-time random walk
on the hierarchical lattice. This arrangment allows us to distinguish the 
physically meaningful (leading order) magnitudes. From this, we analyze some 
results in next section.\

We can choose either representation to obtain the final formulae for
parameters and remainders. Here and in Rodr\'{\i}guez-Romo S. \cite{Su} we 
choose the one to provide new stochastic meaning to renormalizing SUSY field 
theories.\

We claim that this result is the space-time renormalization-group 
trajectory, for the weakly SARW energy interaction studied by Brydges, Evans 
and Imbrie \cite{Ev} provided $\varphi=2$ and $d=4$. In their paper the 
trajectory of a SUSY $\phi^{4}$ was studied (recall that this can be 
understood in terms of intersection of random walks due to Mc Kane, Parisi, 
Sourlas theorem) from a SUSY field-theoretical version of the 
renormalization-group map, on almost the same  hierarchical lattice we are 
studying here. We improve the model by providing exact expressions for 
$\lambda$ and $m$ for each step the renormalization-group is applied in the 
stochastic framework, among others.\

To obtain eq(13) we have introduced an initial mass term $m$, $O(\lambda^2)$ 
(this allows a factorization whose errors are included in the remainder $r'$, 
automatically). We use the Duplantier's hypotesis \cite{Du} and assume all 
divergences of the SUSY $\phi^4$ with ultrametric as coming from the vertices 
or interactions per site of the lattice. This hypothesis has been proved to be 
correct in dimension 2 by means of conformal field theory. Then, a formal 
Taylor series expansion is applied which is analyzed for each particular 
topology in the renormalized field theory (this is done in random walk 
representation) per site of the new lattice. Putting everything together and 
by induction, we obtain the final result.\

We can apply the very same method to study any SUSY $\Phi^n$ model on this 
ultrametric space.\

\section{Renormalized SUSY $\phi^4$ with ultrametric. The stochastic 
approach.}
To start with, we write the physically meaningful (leading order) part of 
eq(13); namely eq(14), eq(15) and eq(16) in parameter space. Let us define the 
following vector
\begin{equation}
{\bf H}=(m,\lambda)
\end{equation}
Here we have approached up to the most probable events (first order in SUSY 
representation). The action of the renormalization-group map (RG) is expressed 
as 
\begin{equation}
{\bf H'}=R({\bf H})=(m',\lambda').
\end{equation}
The fixed points in our theory, $(m^*_1,\lambda^*_1)$ and 
$(m^*_2,\lambda^*_2)$ are as follows.\\

a) The trivial $m^*_1$=$\lambda^*_1$=$0$.\\

b) $\lambda^*_2$=$\frac{L^{2\varphi-d}-1}{\chi}$ ; 
$m^*_2$=$\frac{\gamma_1(L^{2\varphi-d}-1)}{\chi(1-L^{\varphi})}-
\frac{\gamma_2(L^{2\varphi-d}-1)^2}{\chi^2(1-L^{\varphi})}$.\\

The nontrivial fixed point involves a renormalized two-body interaction which
is inverse to the conditional expectation of two two-body interactions that
renormalizes to a two-body interaction inside the contracting ${\it G}_1$ 
cosets ($\chi$) given that the RG  map is applied. Meanwhile the renormalized 
mass in this point is given in terms of two ratios. The first one involves the 
ratio of conditional expectations of one two-body interaction that 
renormalizes to local times ($\gamma_1$) inside a contracting ${\it G}_1$ 
coset and $\chi$. The second ratio involves the conditional expectation of
two two-body interactions that renormalize to local times ($\gamma_2$) inside
a contracting ${\it G}_1$ coset and $\chi^2$. Both; $\lambda^*_2,m^*_2$, are
independent of the scaling factor $L$ for large lattices.\

As we come infinitesimally close to a particular fixed point, 
(called this ${\bf H^*}$), the trajectory is given completely by the single 
matrix $M$ (its eigenvalues and eigenvectors). Namely
\begin{equation}
M_{ij}=\left.\frac{\partial R_i({\bf H})}
{\partial H_i}\right|_{{\bf H}={\bf H}^*}
\end{equation}
From the random walk representation of SUSY $\phi^4$ with ultrametric, up to
the most probable event approach (leading order in SUSY representation), we 
obtain
\begin{equation}
M=
\left(
\begin{array}{cc}
L^{\varphi}  & \gamma_1-2\gamma_2\lambda^* \\
0 & L^{2\varphi-d}-2\chi\lambda^*
\end{array}
\right),
\end{equation}
where $\lambda^*$ can be either $\lambda^*_1$ or $\lambda^*_2$.\

The eigenvalues and eigenvectors of this matrix are as follows.\\

a) $l_1=L^{\varphi}$ with eigenvector $(m,0)$.\\

b) $l_2=L^{2\varphi-d}-2\chi\lambda^*$ with eigenvector
$\left(m, -\frac{L^{\varphi}-2L^{2\varphi-d}+2\chi\lambda^*}
{\gamma_1-2\gamma_2\lambda^*}m\right)$, where $\lambda^*$ can be either 
$\lambda^*_1$ or $\lambda^*_2$.\\

For $L\geq 2$ and $\varphi> 0$; both fixed points $(m^*_1, \lambda^*_1)$
and $(m^*_2, \lambda^*_2)$ are repulsive in the direction of the eigenvector
$(m,0)$, marginal if $\varphi=0$ and attractive if $\varphi<0$. The trivial 
fixed point $(m^*_1, \lambda^*_1)$ is repulsive in the direction of the 
eigenvector 
$\left(m,-\frac{L^{\varphi}-2L^{2\varphi-d}+2\chi}{\gamma_1} m\right)$
provided $\varphi>d/2$, marginal if $\varphi=d/2$ and attractive otherwise. 
Finally, the fixed point $(m^*_2, \lambda^*_2)$ is repulsive in the direction 
of the eigenvector 
$\left(m,-\frac{L^{\varphi}-2L^{2\varphi-d}+2\chi}{\gamma_1} m\right)$ 
provided $d/2>\varphi$, marginal if $d/2=\varphi$ and 
attractive otherwise. This means that the only critical line which forms the 
basin of attraction for both fixed points is given only for $0<\varphi<d/2$ 
and is locally defined by $g_1=0$. Here $g_1$ is the linear scaling field 
associated with the eigenvector $(m,0)$.\

The largest eigenvalue defines the critical exponent $\nu$. In the trivial
fixed point $(m^*_1,\lambda^*_1)$, $\nu=1/\varphi$ provided $d\geq \varphi$.
If $d< \varphi$ than $\nu=\frac{1}{2\varphi-d}$. Here the eigenvalue 
$l_1=L^{\varphi}$$>1$ provided $2\varphi<d$; i.e. this fixed point is 
repulsive in the direction of the eigenvector $(m,0)$ if and only if 
$2\varphi<d$. Although our results are rather general, let us consider the 
Flory's case as an example \cite{Fl}. For $d\geq 5$ this trivial fixed point, 
in the Flory's case, is attractive in the direction of the eigenvector 
$(m,0)$, marginal in dimension four and repulsive otherwise.\

In the fixed point $(m^*_2,\lambda^*_2)$, $\nu=\frac{1}{\varphi}$, provided 
$\beta>-log_L\left(\frac{1+L^{\beta-d}}{2}\right)$. Back to the example we are 
considering here (Flory's case) \cite{Fl}; for $d\geq 5$ this fixed point is 
repulsive, marginal in $d=4$ and attractive otherwise.\

We cannot explain, from this first order approach (the most probable event), 
logarithmic corrections to the end-to-end distance in the critical dimension. 
This is correctly explained, although heuristically, elsewhere \cite{Su}.\

Using the spin representation, we find the following.\\

a) For the trivial fixed point $(m^*_1,\lambda^*_1)$.\\

$\alpha$=$2-d/\varphi$ ; $\beta$=$\frac{2(d-\varphi)}{\varphi}$ ;
$\gamma$=$\frac{4\varphi-3d}{\varphi}$ ; 
$\delta$=$\frac{2\varphi-d}{2d-2\varphi}$ ; 
$\nu$=$\frac{1}{\varphi}$ and finally $\eta$=$2-4\varphi+3d$.\\

b) For the fixed point $(m^*_2, \lambda^*_2)$.\\

$\alpha$=$2-d/\varphi$ ; $\beta$=$\frac{d-log_L(2-L^{2\varphi-d})}{\varphi}$ ; 
$\gamma$=$\frac{2log_L(2-L^{2\varphi-d})-d}{\varphi}$ ; 
$\delta$=$\frac{Log_L(2-L^{2\varphi-d})}{d-log_L(2-L^{2\varphi-d})}$ ; 
$\nu$=$\frac{1}{\varphi}$ and finally $\eta$=$2+d-2Log_L(2-L^{2\varphi-d})$.\\

Besides, if we introduce critically the mass as was done in Brydges et al. 
\cite{Ev} in $d=4$ and $\varphi=2$, the critical exponents look as follows.\\

$\alpha$=$0$ ; $\beta$=$\frac{1}{2}$ ; $\gamma$=$1$ ; $\delta$=$3$ ; 
$\nu$=$\frac{1}{2}$ and finally $\eta$=0.\\

On the other hand, we know that for the SUSY  $\lambda\phi^4$, 
$\beta(\lambda')$=$\mu\frac{\partial \lambda'}{\partial \mu}$, where $\mu$ is 
a parameter with the dimensions of mass; namely $\mu$ is an arbitrary mass 
parameter.\

Since we know the fixed points for the theory in random walk representation; 
these must be the zeros of $\beta(\lambda)$ in the
SUSY $\lambda\phi^4$ representation. Using this criteria we obtain the
following expression for $\beta(\lambda)$;
\begin{equation}
\beta(\lambda)=\frac{\gamma_2}{1-L^{\varphi}}\lambda^2-
\frac{\gamma_1}{1-L^{\varphi}}\lambda+m
\end{equation}
up to a multiplicative constant.\

An interesting pictorial interpretation of the renormalized group equation was
suggested by S. Coleman \cite{Co}. The equation can be viewed as a flow of 
bacteria in a fluid streaming in a one dimensional channel. Here we provide a 
new interpretation of the velocity of the fluid at the point $\lambda$, 
$\beta(\lambda)$, in terms of stochastic events (function of conditional 
expectations of two-body interactions inside contracting ${\it G}_1$ cosets). 
For large lattice $\beta(\lambda)$ is independent of the lattice parameter 
$L$.\

Concretely, $\beta(\lambda)$ (or the velocity  of the fluid at the point
$\lambda$) is written in terms of one two-body and two two-body contributions
to renormalized local times (stochastic approach) or mass (field theory 
approach). The first contribution is $O(\lambda)$ and the second, 
$O(\lambda^2)$.

Let us call $\beta'(\lambda)
=\left(\frac{\partial \beta(\lambda)}{\partial \lambda}\right)_m$, then
\begin{equation}
\beta'(\lambda)=\frac{2\gamma_2\lambda-\gamma_1}{1-L^{\varphi}}
\end{equation}
In the trivial fixed point, $\beta'(\lambda^*_1)> 0$ (infrared stable), 
provided $\varphi> 0$ and $L\geq 2$; besides $\beta'(\lambda^*_1)< 0$ 
(ultraviolet stable), provided $\varphi< 0$ and $L\geq 2$. In the fixed point 
$(m^*_2,\lambda^*_2)$, $\beta'(\lambda^*_2)\geq 0$ (infrared stable), provided 
$\varphi\geq 1/2\left[d+log_L\left(\frac{\gamma_1\chi}{2\gamma_2}+
1\right)\right]$, and $\beta'(\lambda^*_2)< 0$ (ultraviolet stable) otherwise. 
Here we define $ d_H=log_L\left(\frac{\gamma_1\chi}{2\gamma_2}+1\right)$ which 
is given in terms of the ratio for conditional expectations of two-body 
interactions which renormalizes to local time and two-body interactions. From 
this, the following estimates are obtained \\

a) d=4; $\beta'(\lambda^*_2)\leq 0$, provided $d_H\geq 0$.\

b) d=3; $\beta'(\lambda^*_2)\leq 0$, provided $d_H\geq 1/3$.\

b) d=2; $\beta'(\lambda^*_2)\leq 0$, provided $d_H\geq 2/3$.\

d) d=1; $\beta'(\lambda^*_2)\leq 0$, provided $d_H\geq 1$.\

\section{Summary}
Because of the equivalence between the polymer and SAW problems, functional
integration methods were employed in the majority of theoretical approaches to
these problems. It should, however, be remarked that the critical exponents
for the SAW obtained by this method are only meaningful if the spatial
dimensionality $d$ is close to its formal value $d=4$, and it is not yet clear
how to get results for real space in this way. There is another method based 
on the search for a solution to the exact equation for the probability density
of the end-to-end distance of the random walk \cite{Al}. By defining the self-
consistent field explicitly, the density could be found with the help of the
Fokker-Planck equation. In this paper we provide another alternative view 
where the probability density, as a random function of the random walk, is 
proposed.\

Discrete random walks approximate to diffusion processes and many of the 
continuous equations of mathematical physics can be obtained, under suitable
limit conditions, from such walks. Besides we can look at this relation the 
other way around; that is, suppose that the movement of an elementary particle
can be described as a random walk on the lattice, which represents the
medium it traverses, and that the distance between two neighbouring vertices,
though very small, is of a definite size; therefore the continuous equations 
can be considered as merely approximations which may not hold at very small
distances. We show in this paper how the mathematical results are easily 
derived by standard methods. The main interest lies in the interpretation of
the results.\

In our approach the properties of the medium will be described by the lattice 
and the transition probabilities. We obtain $m'$, the ``mass" of the field as observed 
on this particular hierarchical lattice. The lattice is characterized by the 
ultrametric space used to label this.\

We propose to obtain renormalized $n$-body interactions out of a set of 
stochastic diagrams with a fixed totally arbitrary topology.\

Here we would like to stress that the search for a proper mathematical 
foundation of a physical theory does not mean only a concession to the quest
for aesthetic beauty and clarity but is intended to meet an essential physical
requirement. The mathematical control of the theory plays a crucial role to 
allow estimates on the proposed approximations and neglected terms.\

Usually approximations must be introduced which often have the drawback that,
although they can work well, they are uncontrolled: there is no small 
parameter which allows an estimate of the error.\

Explicit mathematical formulae for all the parameters and remainders in the
method can be provided. In sake of brevity we present these elsewhere
\cite{Su}. All of them are expressed in terms of conditional expectations of
events inside contracting ${\it G}_1$ cosets.\

Once a successful theoretical scheme has been found it is conceivable that it
is possible to reformulate its structure in equivalent terms but in different 
forms in order to isolate, in the most convenient way, some of its aspects and
eventually find the road to successive developments.\

Let us remark that we are talking of a particle picture even when we deal with
systems containing many particles or even field systems.\

We hope our method and ideas may help in the proper understanding of the 
association of stochastic processes to the quantum states of a dynamical 
system; i.e. stochastic quantization.\

Summarizing, in this paper we present an heuristic space-time \newline
renormalization-group map, on the space of probabilities, to study SUSY 
$\phi^4$ in random walk representation, on a hierarchical metric space defined 
by a countable, abelian group ${\it G}$ and an ultrametric $\delta$.  We 
present the L\'evy process on $\Lambda_n$ that correspond to the random walk 
representation of SUSY $\phi^4$  which is a configurational measure model 
from the point of view of a stochastic processes. We apply the 
renormalization-group map on the random walk representation and work out 
explicitly the weakly SARW case for double intersecting paths which 
corresponds to SUSY $\phi^4$, as an example. The generalization to SUSY 
$\phi^n$, for any $n$, is straightforward. New conclusions are derived from 
our analysis.\

Our result improves the field-theoretical approach \cite{Ev} by obtaining an 
exact probabilistic formula for the flow of the interaction constant and the 
mass under the map.

\section{Acknowledgments}
This research was partially supported by CONACYT, Ref 4336-E, M\'exico.

\end{document}